\documentclass[12pt]{iopart}

\usepackage{a4wide}
\newcommand{\nm}{\mbox{nm}}   
\newcommand{\um}{\ensuremath{\mu}m}
\usepackage{graphicx}

\begin{document}

\title[Cytosim]{Collective Langevin Dynamics of Flexible Cytoskeletal Fibers}
\author{Francois Nedelec\dag\ and Dietrich Foethke}

\address{European Molecular Biology Laboratory\\
69117 Heidelberg, Germany}
\address{\dag To whom correspondence should be addressed}
\ead{nedelec@embl.de}
\date{May 2, 2007}

\begin{abstract}
We develop a numerical method to simulate mechanical objects in a viscous medium at a scale where inertia is negligible. Fibers, spheres and other voluminous objects are represented with points. Different types of connections are used to link the points together and in this way create composite mechanical structures. The motion of such structures in a Brownian environment is described by a first-order multivariate Langevin equation. We propose a computationally efficient method to integrate the equation, and illustrate the applicability of the method to cytoskeletal modeling with several examples.
\end{abstract}


\maketitle

\section{Introduction}

The internal architecture of living cells relies largely on microscopic fibers, which form the cytoskeleton with their associated proteins. These fibers have remarkable mechanical properties. Microtubules and actin filaments for instance have persistence lengths of $\sim$5~mm and 20~\um, respectively, and can sustain pico-Newtons of force without breaking \cite{Howard2001}. Yet these fibers can also be broken down quickly, because they are formed by the non-covalent assembly of protein monomers. Filament ends can grow or shrink, or even alternate between those two states in a remarkable process called dynamic instability \cite{Desai1997a, Garner:2004aa}.
Structurally, the monomers in microtubules and actin filaments assemble head to tail in a regular manner. On the resulting polar lattices, mechano-enzymes called molecular motors (for example kinesin on microtubules or myosin on actin-filaments) use chemical energy to move directionally \cite{Howard2001} or to organize the filaments in space \cite{Wittmann2001}. Furthermore, specific enzymes control the filaments by regulating nucleation, assembly/disassembly or even by severing the filaments. 

The cytoskeleton is involved in multiple cellular processes such as cytokinesis, motility, polarization and mitosis.
These functions are accomplished by many filaments working together. In this way, a set of dynamic or short-lived filaments may form a stable larger assembly, as exemplified by the mitotic spindle \cite{Wittmann2001}. 
Many of the enzymes involved in the assembly of these structures are part of multi-functional entities \cite{Blaineau:2007aa, Varga2006, Janson:2007aa}.
For example, motors form oligomers that can actively connect filaments together \cite{Wittmann2001}; motors may be able to disassemble filaments \cite{Varga2006}; nucleation can be controlled such that it occurs on existing filaments \cite{Janson2005, Mahoney2006}; crosslinkers may be polarity-specific \cite{Loiodice2005} and motors are sometimes linked to proteins that track the tips of growing microtubules \cite{Janson:2007aa, Ambrose:2005aa, Goshima:2005aa, Morrison:2007aa}.
Generally speaking, modularity allows the cytoskeleton to be reprogrammed, for example at different stages of the cell cycle. It allows cells to reuse the same functional elements to achieve different tasks and multiplies the number of way in which the organization of fibers can be regulated. This modularity is certainly a consequence of the combinatorial exploration operating during natural selection \cite{GerhartKirschner1997}. In any case, the cytoskeleton in addition to fibers contains a kit of activities which can be combined in many ways.

Biological systems are hard to understand, and theory is necessary to approach the non-intuitive aspects \cite{Mogilner2006a}. It is notable that many models in the cytoskeleton field often include the same basic elements (for a recent review on this subject, see \cite{Karsenti2006}). This reflects the inherent modularity of the biological design illustrated briefly in the previous paragraph, and also affects the modeling approach. It implies that it is worthwhile to build a computer simulation to model a few basic elements, if these elements can be combined freely to rapidly model diverse situations. In practice, the elements of the simulation (\textit{eg.} a model of kinesin, or a model of a severing enzyme) can even be implemented, tested and benchmarked by different teams of experts for each aspect of the system. Sharing computer code in this way can in fact be a practical mean to combine the efforts of the community.

Writing a cytoskeletal simulation is likely to be a collective task also because it is a demanding project, involving multiple aspects: 
\textbf{(a)}~chemical reactions that occur inside cells, 
\textbf{(b)}~transport along fibers, for example the motion of molecular motors,
\textbf{(c)}~assembly dynamics of cytoskeletal fibers and
\textbf{(d)}~motion and deformation of fibers. 
Fortunately, numerous algorithms are available for certain of these aspects, in particular for the reaction-diffusion (see \cite{Lemerle:2005aa, Dobrzynski:2007aa}). Transport along fibers can be modeled with advection equations, or with more details of the motion of the motors \cite{Kolomeisky:2007aa}.
The assembly dynamics of fibers has been the subject of much research and cannot be reviewed here (see \cite{Karsenti2006}). 
The deformation of the fibers is a classical mechanical problem (see for example \cite{Landau1986, Feynan1989}). However, the scale of living cells is associated with many specific features. In particular, Brownian motion plays a fundamental role, inertia is negligible \cite{Purcell1977} and the fibers are dynamic: they can lengthen or shorten by self-assembly. As a consequence, the physics of biological fibers is fundamentally distinct from other mechanical systems. In brief, public or commercial codes are not adapted to simulate the cytoskeleton.

The purpose of this paper is to describe a method to calculate the mechanics of an ensemble of connected fibers and other objects, which is the basis of a cytoskeletal simulation such as cyto{\bf sim}. The physics of such system is described by a Langevin equation (for an introduction, see \cite{Lemons1949}) that recreates the Brownian motion of the fibers and includes bending elasticity, fiber-fiber interactions and external force-fields.
Following earlier work \cite{Deutsch1988, Bourdieu1995}, we use constraints in order to maintain the length of the fibers. This is an alternative to methods in which potentials are used to represent the longitudinal stiffness of fibers. We extend this approach by introducing an implicit integration scheme.
Our method was first used to simulate the effects of motor complexes on two radial arrays of microtubules (asters) \cite{Nedelec2002}, and more recently the assembly of anti-parallel microtubule arrays in \textit{S.~pombe} \cite{Janson:2007aa} and the positioning of the spindle in the \textit{C.~elegans} embryo \cite{Kozlowski:2007aa}.
A major aim of these simulations was to reconstitute the system's operation \textit{in silico}, from established physical principles. This offers two major advantages: i) the assumptions of the model are well defined and can always be modified; ii) any property of the system can be measured easily. This facilitates further investigations. For example we could systematically simplify the model in order to identify a minimal set of working properties \cite{Janson:2007aa}.  In addition, we could identify the parameter range under which the system can operate \cite{Kozlowski:2007aa}. However, for these results to be valid, the systems operation needs to be reproduced correctly at the first place! To maximize the chances of success, it is desirable to reconstitute the mechanics in a physically sensible and accurate way. One may otherwise derive conclusions which do not apply to the real system.

In this paper, we focus on the mechanical aspects of the fibers, and explore the numerical resolution of the associated equations. 
We first describe objects that in addition to fibers are useful for simulating different cellular skeletons. We then present the equation of motion and discuss its numerical integration. We examine the numerical stability of the resulting method and discuss how it affects the simulation speed. Finally, we discuss how other aspects of the cytoskeleton can be added to extend the mechanical calculation.

\section{Objects}

\begin{figure}[p]
\centering
 \includegraphics[scale=0.8]{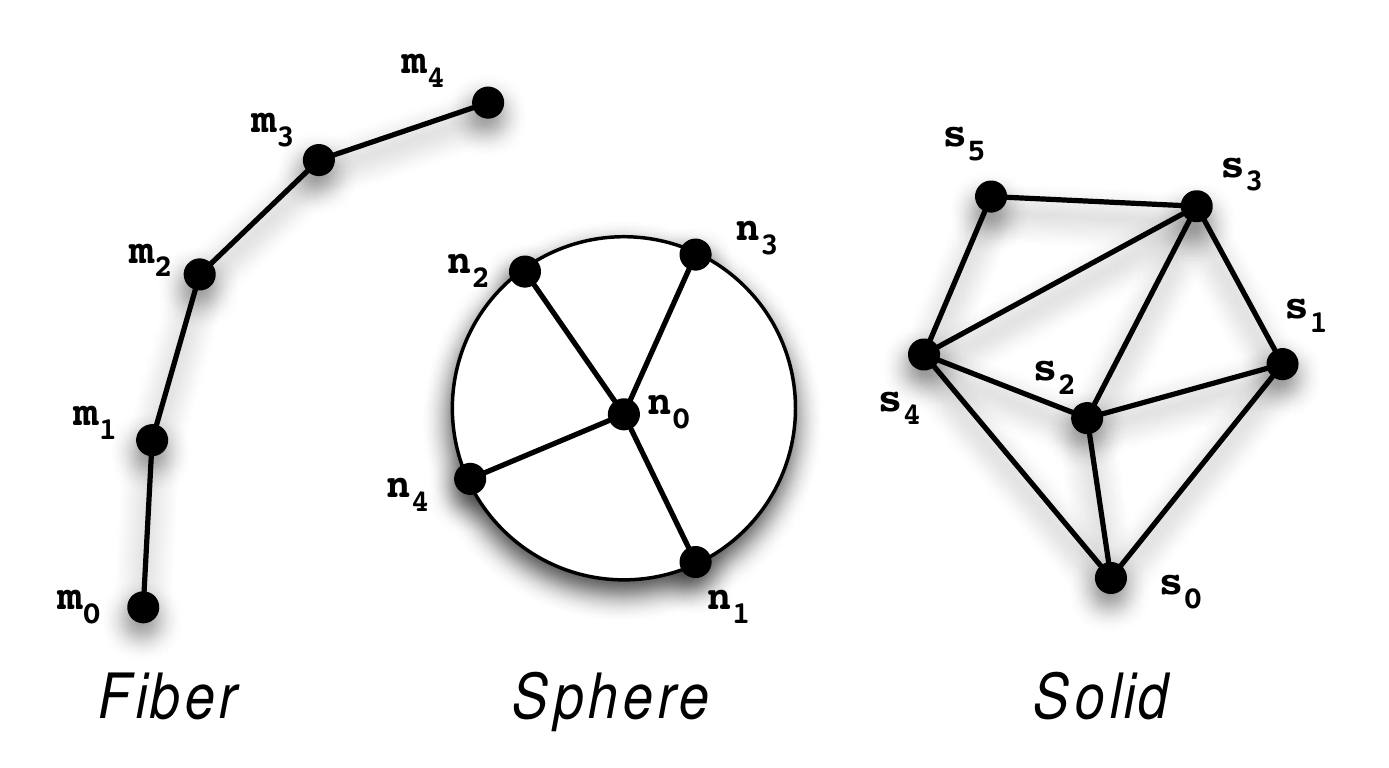}
 \caption{\label{fig:objects} \textbf{Elementary Objects.} All objects in the simulation are described by points. The points can move in the viscous medium, but the relative distances between certain points are conserved (lines). {\bf Left:} A \textit{fiber} is modeled as an equidistributed string of points. {\bf Center:} A \textit{sphere} is composed of a central point and peripheral points, located a distance $r$ from the center. The peripheral points can move on the surface, as if they were in a viscous membrane. {\bf Right:} A \textit{solid} is a set of points that behaves like a solid body. Its shape and size are constants.}
 \end{figure}

\begin{figure}[p]
\centering
\includegraphics[scale=0.4]{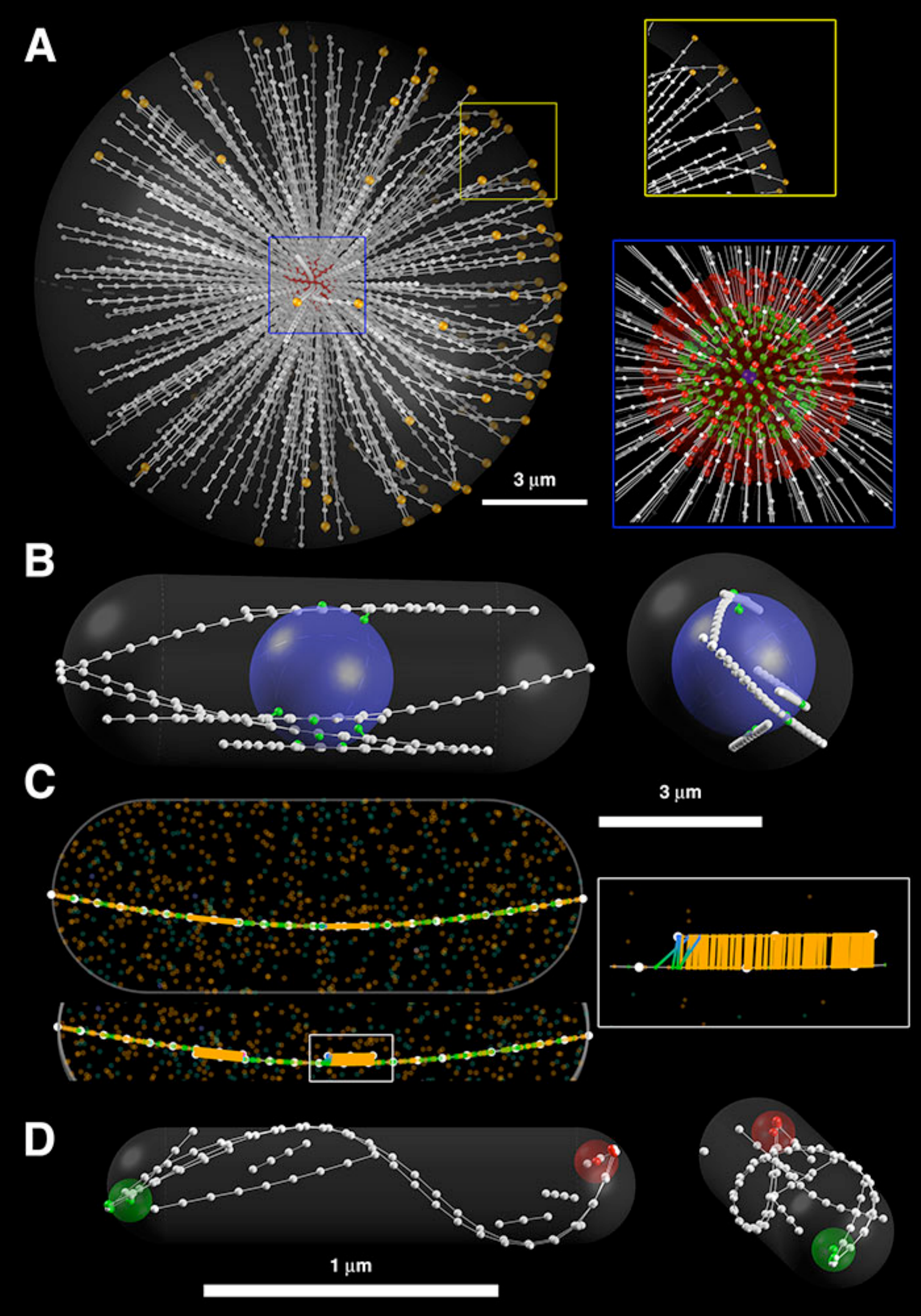}
 \caption{\label{fig:examples} \textbf{Some problems studied with cyto{\bf sim}.} In all the images, \textit{fibers} are indicated in white, along with their model-points. ({\bf A}) An aster is constructed by assembling fibers radially around a solid \cite{Nedelec2002, Kozlowski:2007aa}. Right, top: interactions of microtubules with the cell cortex. Right, bottom: the \textit{solid} is made of a central point (blue) surrounded by two concentric layers of peripheral points (green and red). Only the central point is associated with a viscous drag ($a_i>0$). The other points are used to attach fibers: the minus-end to one green point, and a distal position on the fiber to one red point.   
Using a similar simulation with two asters linked by a \textit{solid} spindle, we proposed an original model describing the 3D motions of the spindle in the first cell division of the \textit{C. elegans} embryo \cite{Kozlowski:2007aa}.
 ({\bf B}) Microtubules in interphase fission yeast  and the nucleus, represented by a \textit{sphere} (blue/green). This can be used to study the role of mechanics in regulating the dynamics and organization of microtubules. 
 ({\bf C}) Self-assembly of interphase microtubules arrays in fission yeast. The simulation contains no steric interaction between the fibers, and they overlap freely. In the display, however, the fibers are shifted in order to visualize the bridging complexes (bottom and right). Using this simulation, we could identify a minimal `recipe' to make stable bundles from dynamic microtubules. This recipe describes how cross-linking, nucleating and motor activities can be associated to obtain the result observed \textit{in vivo}. 
 ({\bf D}) Self-segregation of plasmids in prokaryotes. Actin-like filaments are simulated, together with two \textit{solids}, representing the plasmids \cite{Moeller-Jensen2003}. The efficiency of the segregation is recapitulated in the simulation, and can therefore be analyzed. }
 \end{figure}

More accurate mechanics can be achieved if we introduce two new objects in addition to \textit{fibers}: spherical sets of points (\textit{spheres}) and non-deformable sets of points (\textit{solids}). 
These objects are also described with points but have different morphologies (see fig.~\ref{fig:objects}). The mechanical properties are also distinct. While \textit{fibers} may bend, the \textit{solids} do not deform. The \textit{spheres} can represent spherical viscous membranes such as vesicles.
Any number of objects can be combined in various ways to build complex cytoskeletons.
For example, to simulate interacting microtubule asters \cite{Nedelec2002}, \textit{fibers} were positioned around a \textit{solid} using static links (see fig.~\ref{fig:examples}A). The solid represented in this case the organelle (called the centrosome) which in the cell generates microtubules in a radial fashion. \textit{In vivo} as well as in the simulation, the resulting structure is radially symmetric, and the fibers have their ends mechanically joined together.
Two such asters were further connected by another \textit{solid}, to model the positioning of the mitotic spindle in \textit{C.~elegans} \cite{Kozlowski:2007aa}. In this case, the additional solid represented the pole-to-pole mechanical connection achieved by the mitotic spindle.
To simulate nuclear positioning in \textit{S.~pombe},  \textit{fibers} (microtubules) were attached to a \textit{sphere}, and the ensemble was confined in a cylindrical volume (see fig.~\ref{fig:examples}B). 
The \textit{fibers} and the \textit{sphere} represented microtubules and the cell nucleus, which are attached also in the real cell.
To model the formation of anti-parallel microtubule arrays in \textit{S.~pombe} \cite{Janson:2007aa}, \textit{fibers} where connected by motors and other crosslinkers (see fig.~\ref{fig:examples}C). 
Using \textit{fibers} and \textit{solids}, it is also possible to model the segregation of parM plasmids in \textit{E.~coli}  (see fig.~\ref{fig:examples}D), a process which depends on actin-like filaments \cite{Moeller-Jensen2003}.
The  objects can naturally be combined in many more ways than illustrated here.
This enables diverse cellular mechanics to be reproduced, and consequently widens the application scope of the method.
This freedom is intimately linked to the structure of the master equation that will be examined below, and to the way it is integrated numerically.

\section{Constrained Langevin Dynamics}

In the simulation, fibers and other objects are described by points. 
The coordinates of the points are collected in a vector $\mathbf{x}$ of size $Nd$, for a system of $N$ points in dimension $d$.
Following Langevin (for a simple introduction, see \cite{Lemons1949}) the equation of motion reads:
\begin{equation}
\label{eq:dynamics}
d\mathbf{x} = \mu\, F(\mathbf{x},t) \, dt + dB(t)
\end{equation}

$F(\mathbf{x},t)$ of size $Nd$ contains the forces acting on the points at time $t$.  It includes object-specific forces such as bending elasticity, and all the links between different objects. $dB(t)$ of size $Nd$ summarizes the random molecular collisions leading to Brownian motions; it is a stochastic non-differentiable function of time. 
The matrix $\mu$ contains the mobility coefficients of the object-points, which will be defined later for each object. 

\begin{figure}[p]
\centering
\includegraphics*[scale=0.8]{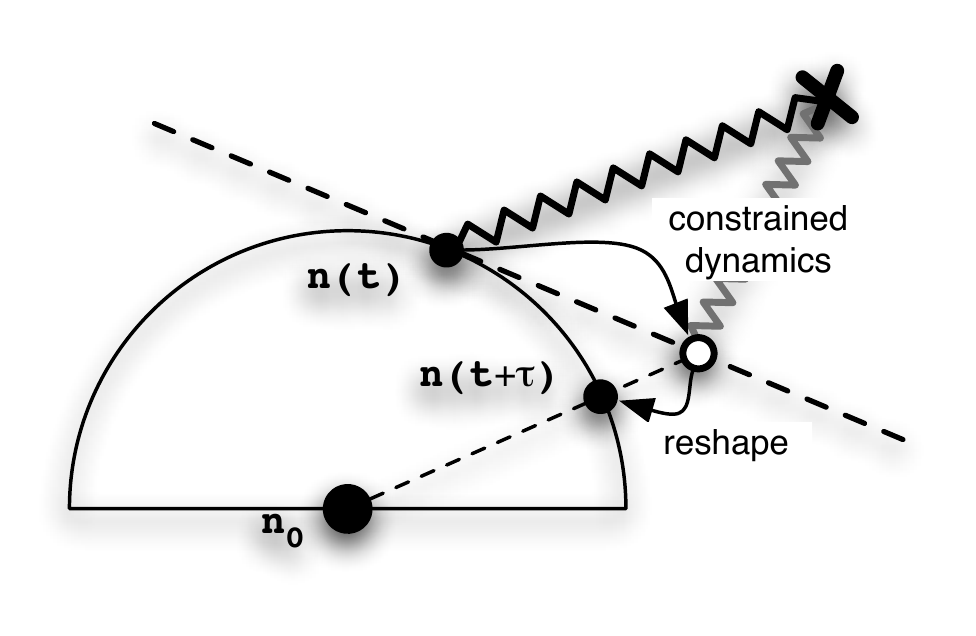}
 \caption{\label{fig:projection} \textbf{Dynamics with Constraints.} The principle of the algorithm is illustrated here for a point $n_i$ constrained to stay at a fixed distance from $n_0$. The point is first moved on the tangent to the circle (this is the plane associated with the constraint) using an implicit integration scheme. The constraint is then re-established exactly by projecting on the sphere. We call this last operation `reshaping an object'. }
 \end{figure}

In addition, certain distances between points inside the objects ($|a_i-a_j| = \lambda_{ij}$) must be conserved during the motion. 
To satisfy these constraints, we perform a step of the dynamics in a subspace tangent to the manifold defined by the constraints, and project the result on the manifold. 
The procedure can be explained simply for a point $n$ constrained to move at a distance $r$ from a fixed position $n_0$ (see fig.~\ref{fig:projection}).
To calculate the motion of $n$, we first write its dynamics in the plane tangent to the sphere at the current position (this is the plane allowed by the constraint $|n - n_0| = r$). The restricted dynamics is integrated implicitly, and the result projected on the sphere to restore the constraint exactly. This approach can be generalized as described next.

\section{Numerical integration}
\label{integration}

From an initial configuration, the system is calculated by discrete time steps $\tau$ (see \cite{NumericalRecipes} for a general discussion on numerical integration). To calculate $\mathbf{x}_{t+\tau}$ from $\mathbf{x}_t$, the equation (\ref{eq:dynamics}) is integrated implicitly. We will discuss the advantages of using an implicit rather than an explicit integration in section \ref{stability}, and concentrate here on the practical issues.
For an implicit integration, we need to express $F(\mathbf{x},t)$ linearly as $A_t\,\mathbf{x} + G_t$, where the square matrix $A_t$ contains the stiffness coefficients associated with the interactions, and the vector $G_t$ contains the constant forces.
This linearization is obtained by summing over all the interactions present at time $t$ (see fig.~\ref{fig:interactions}).
In our simulations, many of the interactions were modeled as harmonic potentials for simplicity, and are therefore already linear. Non-linear interactions simply need to be linearized at this point. In particular, the linearization of the constraints leads to an orthogonal projection $P(\mathbf{x})$, which will be defined later for each object.
To obtain a finite difference scheme for the interval $[t,\,t+\tau]$, $P$ and $A$ are used at time $t$, but $\mathbf{x}$ is used at $t+\tau$ (using $x_{t+\tau}$ instead of $x_t$ is the basis of implicit integration):
$$ \nonumber
\mathbf{x}_{t+\tau}-\mathbf{x}_t  \;=\;   P_t  \left[\; \tau\mu(A_t\,\mathbf{x}_{t+\tau} + G_t) + \delta B_t \;\right] ,
$$
leading to a system of linear equations:
\begin{equation}
\label{eq:integration}\\
\left[\,  I - \tau P_t \mu A_t \,\right]\, (\mathbf{x}_{t+\tau} - \mathbf{x}_t) \;=\;   P_t \left[ \;  \tau\mu ( A_t \, \mathbf{x}_t + G_t) + \delta B_t \;\right],
\end{equation}

where $A_t = A(t)$, $G_t = G(t)$, $P_t=P(x_t)$. 
The ``simulated Brownian'' $\delta B_t = \int_t^{t+\tau} dB$ is a vector $\{ \beta_i \, \theta_{t, i} \}_{i \in [1,Nd]}$, where $\theta_{t,i} \sim \emph{N(0,1)}$ are $Nd$ independent normally distributed numbers (derived from uniformly distributed pseudo-random numbers \cite{NumericalRecipes}). The factors $\beta_i \sim \tau^{1/2}$ represent the magnitude of the Brownian motion during a lapse of time $\tau$. We will see later how they are obtained by calibrating the diffusive motion for the objects. 
The equation can be solved to obtain $\mathbf{x}_{t+\tau}$, since both the right-hand side and the matrix $\left[  I - \tau P_t \mu A_t \right]$ are known.
It would be inefficient to invert the matrix, because the system is sparse (it only has few non-zero coefficients). 
This is true of matrix $A_t$, as long as objects are only connected to few others. 
This is also true of $P_t$ which is block-diagonal: it has one block for each object on the diagonal, but the rest of the coefficients are null. This is because the constraints never involve points from different objects, and the projection can thus be done independently for each object.
In this situation, it is advantageous to solve the linear system using an iterative method \cite{NumericalRecipes}. Different iterative solvers are adapted to different matrices. Because $P_t A_t$ is non-symmetric, we have used the biconjugate gradient stabilized (http://www.netlib.org). This method iteratively converges toward the solution of the linear system, and can be stopped when the difference with the exact solution is below a certain threshold.
We set this threshold to $\psi \, min(\beta_i)$, with $\psi = 1/10$. In this way, the numerical error on $\mathbf{x}$ remains below $10\%$ of the Brownian motion, and the approximate solution of (\ref{eq:integration}) is practically indistinguishable from the real one. 
In practice, it is wise to systematically vary $\psi$ and $\tau$ for each application to check the convergence of the method. It is easy to verify, for example, that more stringent values of $\psi$ produce the same results.

Finally, since equation (\ref{eq:integration}) is obtained by linearization, an additional correction is necessary to re-establish the constraints. The result of equation (\ref{eq:integration}) is projected back on the manifold associated with the constraints \cite{Nedelec2002}. This introduces corrections which are second-order in $\tau$. In the following sections,  we will call this procedure `reshaping' the objects. We now survey how \textit{fibers}, \textit{spheres} and \textit{solids} are represented in space, their mobility coefficients, projection operators and `reshaping' procedure. The interactions between objects (which contribute to $A_t$ and $G_t$) will be described subsequently.

%%%%%%%%%%%%%%%%%% OBJECTS 

\section{Linear set of points (fiber)}
\label{fiber}
\begin{figure}[p]
\centering
\includegraphics[scale=0.8]{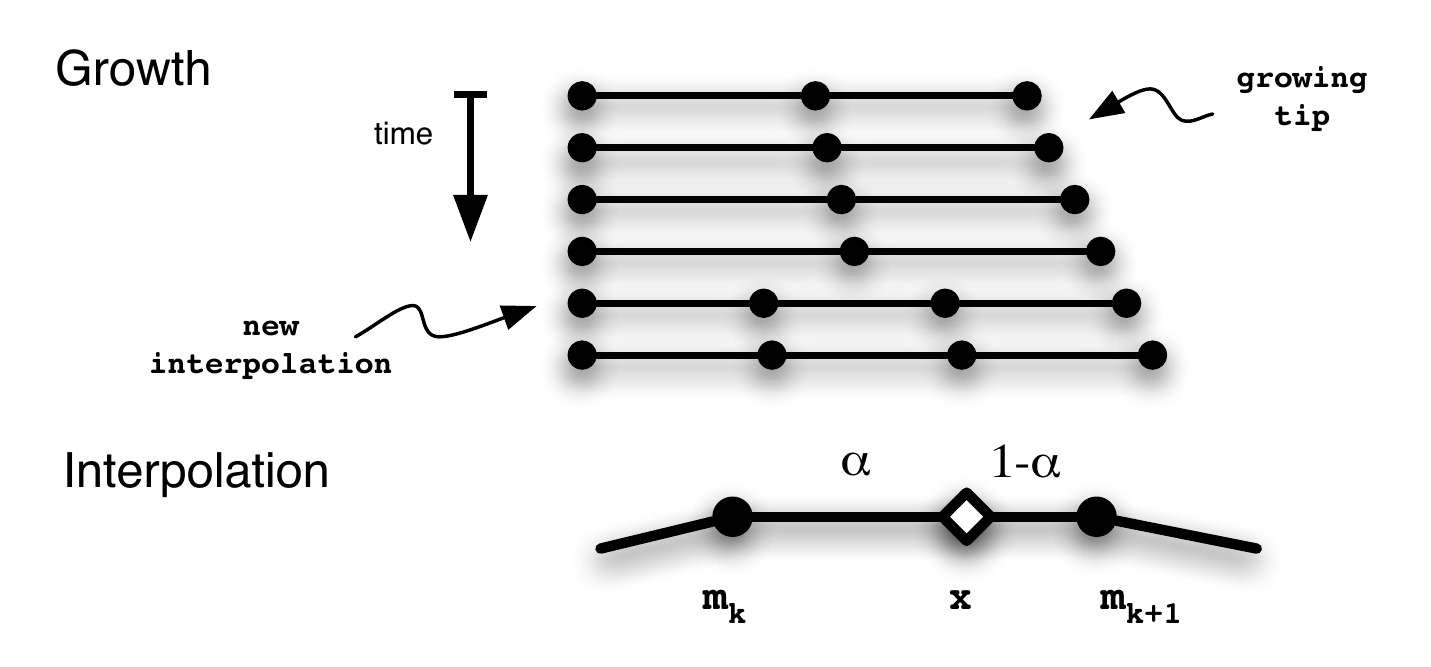}
 \caption{\label{fig:fibers} \textbf{Dynamic Fibers.}  \textbf{Top:} The model-points of a \textit{fiber} are updated when the tips grow, but they are always equally distributed over the fiber. Points are added or removed as necessary to ensure an optimal coverage (see sec. \ref{fiber}). \textbf{Bottom:} An intermediate position $x$ along the fiber is interpolated from the model-points located on each side: $x=(1-\alpha) m_k + \alpha m_{k+1}$ (see sec. \ref{fiber}). }
 \end{figure}

Fibers are modeled as infinitely thin linear objects behaving like elastic, non extensible rods \cite{Nedelec2002}. Each fiber is represented by $p+1$ equidistant model-points $m_i$, for $i \in [0,p]$, separated by a distance $L/p$. A fiber is polar: $m_0$ is the minus-end and $m_p$ the plus-end. The number of segments $p$ is adjusted as a function of the total length $L$ of the fiber. Points are added or removed, in order to always minimize $ | \rho - L/p | $, for each fiber as it grows or shrinks (see fig.~\ref{fig:fibers}). The desired segment length $\rho$ is a parameter  affecting the precision of the simulation. To set $\rho$, one may run a representative case with various values (for microtubules, $\rho < 0.5\; \um$ is usually appropriate). 

\label{fiber:interpolation}
It is often necessary to interpolate between the model-points, when for example calculating the position $x$ of a molecule attached to the fiber.
If $m_k$ and $m_{k+1}$ are the model-points on each side of $x$, we use $x=(1-\alpha) m_k + \alpha m_{k+1}$. The interpolation coefficient $\alpha \in [0,1]$ is calculated from the known relative positions of the three points along the fiber: $\alpha = | m_k x | / | m_k m_{k+1} |$. 
The model-points are themselves updated using this interpolation procedure at every time-step if the length of the fiber has changed (see fig.~\ref{fig:fibers}). 

\subsection{Bending elasticity}
\label{elasticity}

\begin{figure}[p]
\centering
 \includegraphics[scale=0.8]{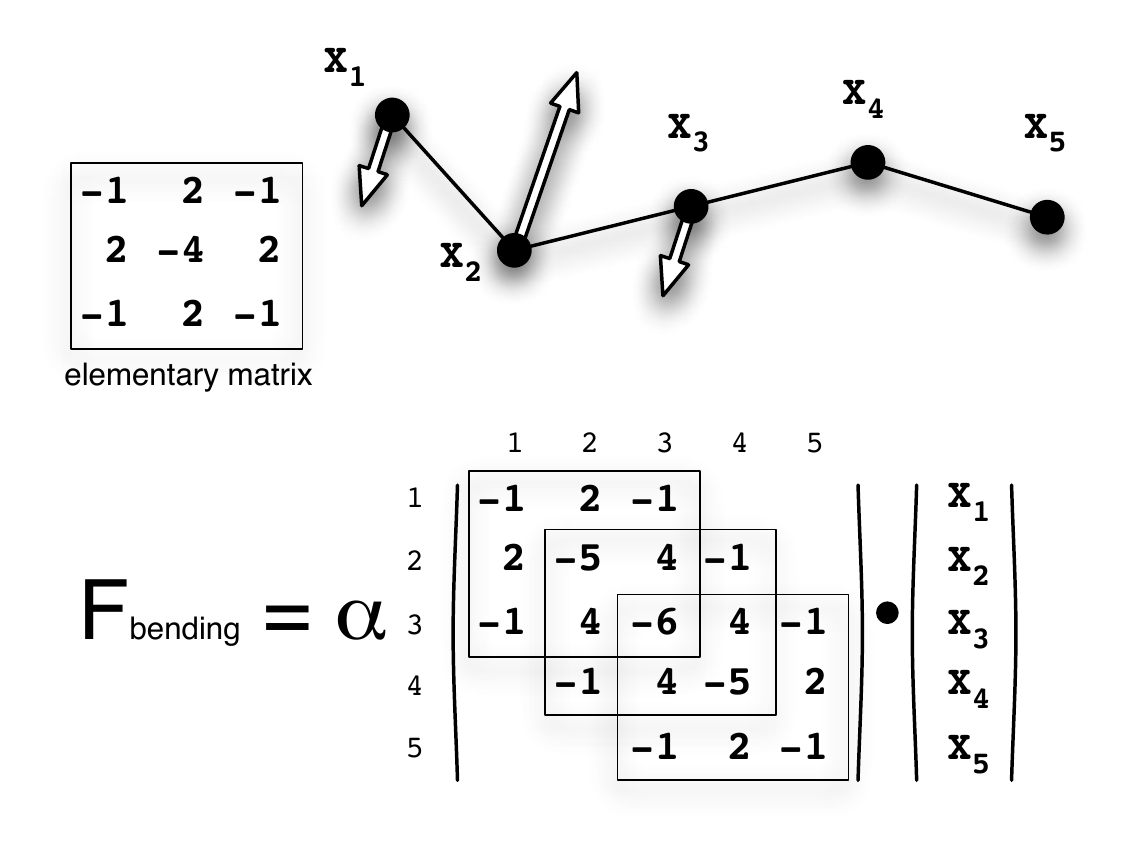}
 \caption{\label{fig:elasticity} \textbf{Matrix elements associated with bending elasticity.} The stiffness matrix $A_t$ contains the bending elasticity of fibers. The contributions are obtained by adding a $3\times 3$ elementary matrix for each consecutive triplets of points (see section \ref{elasticity}). The sum of all rows and columns is zero, since the matrix should only generate an internal torque. The forces associated with the first triplet (points $x_1$, $x_2$ and $x_3$) are depicted. The resulting matrix for 5 points is also shown, and the generalization is straightforward. 
For any fiber, the result is a symmetric banded matrix multiplied by a scalar $\alpha$ that depends on the bending elasticity modulus and on the distance between the points.}
 \end{figure}

\begin{figure}[p]
\centering
 \includegraphics[scale=0.8]{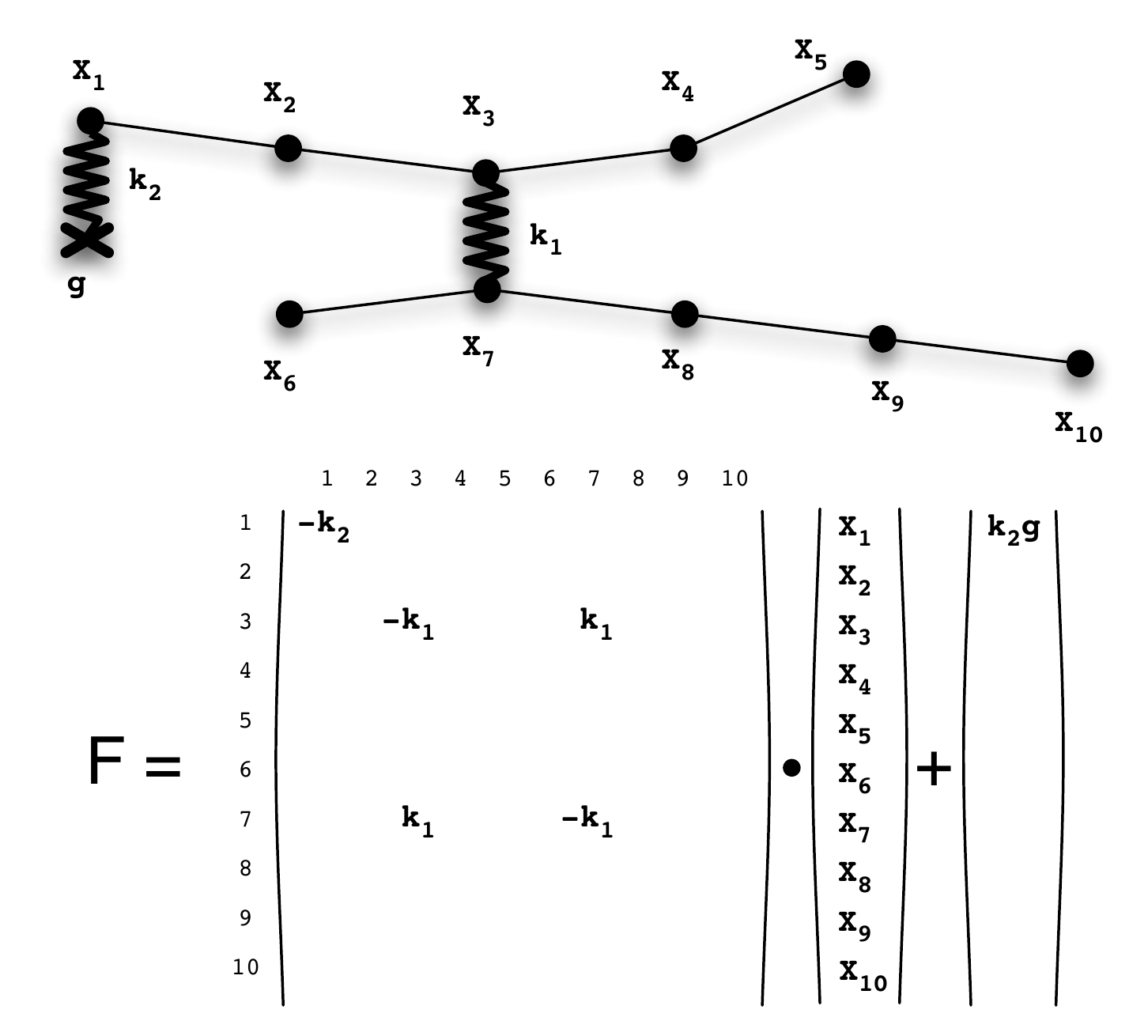}
 \caption{\label{fig:interactions} \textbf{Stiffness matrix and force vector.} The stiffness matrix $A_t$ and the force vector $G_t$ in equation \ref{eq:integration} are set by considering all the interactions present at time $t$. For each interaction, the appropriate formula (sec. \ref{interactions}) is first expanded algebraically. The factors associated with the coordinates of the points are added to $A$, and the coefficients which are independent of the coordinates are added to $G$. At the end of the procedure, one obtains a (sparse) symmetric matrix $A$ and a vector $G$ that provide the forces on the points $F = A\, x + G$. 
Here we illustrate how a connection of stiffness $k_1$ (sec. \ref{connections}) contribute to factors $k_1$ and $-k_1$ at the rows and columns of $A$ corresponding to the points connected. For a connection to a fixed position $g$ (sec. \ref{immobilization}), a stiffness coefficient $-k_2$ is added in $A$, while $k_2 g$ is added in $G$.  
In this example, the connections are attached exactly to points of the system, but this is not always the case. Section \ref{interactions} explains the general proceduce. In addition, the matrix represented here corresponds to a 1D system. It needs to be duplicated for a 2D simulation, and triplicated in 3D (sec. \ref{force:interpolation}). }
 \end{figure}

Fibers can bend under external forces and resist these forces elastically. The standard formula for bending elasticity  \cite{Landau1986} can be applied to strings of points. For any set of three consecutive points  $m_k$, $k \in \{i-1; \; i; \; i+1\}$, we approximate it linearly as a triplet of forces $\{-F; \; 2F; \;-F\}$. Each triplet corresponds to the torque generated between two consecutive segments (see fig.~\ref{fig:elasticity}). Furthermore, we have $F = \alpha (m_{i-1} - 2m_i+ m_{i+1} )$, with $\alpha=\kappa (p/L)^3$, where $\kappa$ is the bending modulus of the fiber, and $L/n$ the length of each segment. The result was verified by comparing the buckling threshold in the simulation with Euler's formula $\pi^2 \kappa/L^2$. The procedure is appropriate if $\rho$ is such that the angles between consecutive segments remain small during the simulation (not shown). Physically, the forces are isotropic, \textit{i.e.} they can be written as a reduced matrix of size $p \times p$ (and not $pd \times pd$), obtained by adding several times the $3\times 3$ matrix $E = - (1,\, -2,\,  1) \otimes (1,\, -2,\, 1)$ ($\otimes$ is the tensor product).
The final result is simple because points are distributed regularly over the length of the fiber (see fig.~\ref{fig:elasticity}).

\subsection{Mobility}

\label{fiber:mobility}
The motion of an object at low Reynolds number is characterized by a mobility. This is defined by factors which link speed and force (speed = mobility $\times$ force). These factors depend on the size and shape of the object, and on the viscosity $\eta$ of the surrounding fluid. 
For instance a straight cylinder has two mobility factors, because it is twofold easier to move in the longitudinal direction than in a transverse direction. This anisotropy could not be implemented simply, because fibers in the simulation may bend and adopt arbitrary shapes. An exact calculation would require finding the hydrodynamic interactions between all the points in the system. This can be done in the future, but for simplicity, we have so far used the averaged mobility of a straight rod of length $L$ and diameter $\delta$: $ \mu = { \log( L_h / \delta ) }/{3 \pi \eta L}$ \cite{Berg1993}. The logarithmic term is an effective hydrodynamic correction on the scale $L_h$, which is either the length of the fiber, or a hydrodynamic cut-off, whatever is smallest. We derive a single mobility factors for the $p+1$ points representing a fiber: $\mu_p = (p+1) \, \mu$.

\subsection{Projector associated with the constraints}

In this section, we calculate the projection $P$ derived from the constraint that the length of the fiber should remain constant during the resolution of equation (\ref{eq:dynamics}).
For each fiber, the coordinates of the $p+1$ model-points $m_k$ are stored in a vector of dimension $(p+1)d$ (for $d=3$, $\{x_0,x_1,x_2\}$ correspond to $m_0$, and $\{x_3,x_4,x_5\}$ to $m_1$, etc).  
The motions of these points are determined by \textit{external} forces $\mathbf{f} = \{ f_k \}$, and additionally by {\it internal} forces $\hat{\mathbf{f}} = \{ \hat{f}_k \}$. The speeds resulting from $\hat{\mathbf{f}}+\mathbf{f}$ should be compatible with the constraints $C_k = (m_{k+1}- m_k)^2 - (L/p)^2 = 0$ for $k \in [0,p[$. 
To calculate $\hat{\mathbf{f}}$ from $\mathbf{f}$, we first define the $p \times d(p+1)$ Jacobian matrix $J_{ij} = \partial C_i / \partial x_j $. In 3D, it reads:
$$
J\! =\! 2 \! \left(\begin{array}{ccccccccccc}
\!x_0\!-\!x_3 & x_1\!-\!x_4 & x_2\!-\!x_5 & x_3\!-\!x_0 & x_4\!-\!x_1 & x_5\!-\!x_2  & 0 & 0 & 0& \!\! \cdots\! \\ 
0 & 0 & 0 & x_3\!-\!x_6 & x_4\!-\!x_7 & x_5\!-\!x_8 & x_6\!-\!x_3 & x_7\!-\!x_4 & x_8\!-\!x_5 & \!\! \cdots\! \\ 
& & & & \vdots  & & & & & \!\!\ddots\!
 \end{array}\right)
$$

Because the mobility coefficients are the same for all the points ($\mu_p$, see sec. \ref{fiber:mobility}), the speed of the points is $\mathbf{v} = \mu_p ( \mathbf{f}+\hat{\mathbf{f}})$. This motion maintains the constraints if $J \, \mathbf{v} = 0$. 
Therefore $\hat{\mathbf{f}}$ must be such that $J(\mathbf{f}+\hat{\mathbf{f}}) = 0$. Furthermore, internal forces should not contribute to global motion or rotation of the object. This imposes that their work should be null for any motion compatible with the constraints: $\hat{\mathbf{f}} \cdot \mathbf{u} = 0$ for any $\mathbf{u}$ such that $J\, \mathbf{u} = 0$. This implies that $\hat{\mathbf{f}} = J^t \mathbf{\lambda}$, where $\mathbf{\lambda}$ is a vector of size $p$ (the Lagrange multipliers). We derive $J(\mathbf{f} + J^t \mathbf{\lambda}) = 0$, and since $JJ^t$ of size $p\times p$ is non-singular, $\mathbf{\lambda} = - (JJ^t)^{-1}J\, \mathbf{f}$, and finally $\hat{\mathbf{f}} = - J^t( JJ^t) ^ { -1} J\, \mathbf{f}$. This shows that the total force can be obtained linearly as $\mathbf{f} + \hat{\mathbf{f}} = P \, \mathbf{f}$, with $P = I - J^t(JJ^t)^{-1}J$. From this result, it is clear that $P$ is an orthogonal projection ($P$ is symmetric and idempotent $PP=P$). Notice that $JJ^t$ is banded symmetric, and therefore easy to invert, which means that $P$ can be computed fast. $P$ (which depends solely on $\mathbf{x}$) is one block of the operator $P_t$ used in equation (\ref{eq:integration}).

Fibers are `reshaped' to restore the constraints exactly after the model-points have been moved. This is done sequentially for $k \in [0,p]$, by moving the points $m_0...m_k$ in the direction of $m_{k+1} - m_k$ and $m_{k+1}...m_p$ in the opposite direction, to restore $|m_{k+1}-m_k| = L/p$ while conserving the center of gravity of the fiber.

\subsection{Brownian motion}

To simulate Brownian motion, a term $\delta B_t$ is attributed to each fiber coordinate $x_t$ (equation \ref{eq:integration}). This term is most simply calibrated by considering diffusion in the absence of bending or external forces ($A=0$, $G=0$). 
If we first assume $P_t=I$ in equation (\ref{eq:integration}), we get $x_{t+h}-x_t =  \delta B_t$. To produce a pure diffusion with a coefficient $D$, one needs:
$$
\langle x_{t+\tau}-x_t \rangle      = 0 \qquad
\langle \, {(x_{t+\tau}-x_t)}^2\rangle   = 2 \, D \,  \tau
$$
This holds true if $\delta B_t$ is normally distributed, of mean zero and variance $2 D \tau$. We can use $\delta B_t = \beta \theta$, where $\theta \sim \emph{N(0,1)}$ is a random number generated for each time step, and $\beta = \sqrt{2 D \tau}$, as mentioned in section \ref{integration}. From Einstein's relation, we set $D=\mu_p k_B T$, where $\mu_p$ is the mobility,  $k_B$ the Boltzmann constant, and $T$ the absolute temperature.
For a fiber with $p+1$ points, we use $(p+1)d$ random numbers, independent and all normally distributed of variance $\beta^2$. 
Projecting these numbers with $P$ produces the appropriate diffusion for the fiber, as well as thermally-driven deformations.
 For example, the translation $x$ of the center of gravity depends on the sum of all the terms in $\delta B$ corresponding to the fiber, leading to a diffusion $D= \mu k_BT$ (with $\mu$ and not $\mu_p$).

\section{Spherical set of points (sphere)}

To simulate the nucleus of \textit{S.~pombe} and attach microtubules on its surface (see fig.~\ref{fig:examples}B), we implemented a `spherical set of points' of radius $r$. Such object is composed of a point $n_0$ in the center, and $q$ additional points $n_i$ on the periphery. If we define $r_k = n_k - n_0$, the constraints are $|r_k|=r$.  A \textit{sphere} moves as a rigid body, and the peripheral points behave as if they were embedded in a viscous surface (see fig.~\ref{fig:objects}). If $f_k$ is the force applied at point $k$, the motion of the set reads:
\begin{eqnarray}
\label{eq:nucleus}
\begin{array}{rcl}
dn_o & = & \mu^T  F \, dt + dB^T \vspace{2mm}  \\
dr_k & = & \left(  \mu^R M \, dt  +  dB^R \right) \times r_k  + P_k \, \left( \mu^S \,f_k \, dt + dB_k^S \right)
\end{array}
\end{eqnarray}
where $F=\sum_{i=0}^q f_i$ is the total force on the sphere, $M=  \sum_{i=1}^q { r_i \times f_i }\;$ is the total torque calculated from the center, and where
$$P_k = \mbox{I} - \frac{r_k \otimes r_k}{r_k^2}$$
 is the projection on the plane tangent to the sphere in $r_k$. $dB^R$, $dB^T$ and $dB_k^S$ are the Brownian terms. 
 Note that these equations would not describe a set of peripheral points articulated around a central node. For example, the motion of the center $n_0$ depends on the sum of all the forces applied to the object, and not only on the force applied in $n_0$. This in fact corresponds to a sphere with points on its surface.
 To keep track of the orientation of the sphere, we also included three reference points $\tilde{n}_k$ on the surface, which form with $n_0$ a reference frame associated to the sphere. The motion of these reference points is entirely determined by the total torque on the sphere: $ d\tilde{r}_k =  \left(  \mu^R M \, dt  +  dB^R \right) \times \tilde{r}_k $, where as before $\tilde{r}_k = \tilde{n}_k - n_0$.
When the object needs to be `reshaped', the peripheral points are simply projected on the surface ($n_0$ is not moved).

\subsection{Mobility and Brownian Motion}

 The equations involve three mobility factors: the translation and rotational mobility of the sphere $\mu^T$ and $\mu^R$, and the mobility of the points in the surface $\mu^S$.
Stokes' law can be used to set $\mu^T$ and $\mu^R$, if the sphere is surrounded by a large volume of fluid. The mobility coefficients for the points in the surface can also be calculated \cite{Saffman1975}.
As described above, points undergo three different types of motion, and a random number $\delta B_t$ in equation (\ref{eq:integration}) is associated with each of these motions. The parameters are calculated by considering diffusion in the absence of other forces ($A=0$ and $G=0$). 
For the translational diffusion of the sphere, the result from equation (\ref{eq:nucleus}) is obtained as previously for the fiber: $\beta^T=\sqrt{2 \mu^T \tau k_BT}$. 
Rotational diffusion is calibrated using equation (\ref{eq:nucleus}).
If $r_t$ is fixed on the surface, we get $r_{t+\tau}-r_t =  \delta B^R_t \times r_t$.
This should be a rotational diffusion of a point on a sphere:
$$
\langle r_{t+\tau}-r_t \rangle      = 0, \qquad 
\langle {(r_{t+\tau}-r_t)}^2\rangle  = 4\, k_B T \mu^R  r^2 \tau. \nonumber
$$
Since $| r_t | = r$, we can use for $\delta B^R_t$ a random vector with $d$ independent components of mean zero and variance ${2 \tau\mu^R k_BT}/{  r^2}$.
A peripheral point $r_t$ also diffuses on the surface, which in equation (\ref{eq:nucleus}) is described by $r_{t+\tau}-r_t =  P_k \; \delta B^S_{k,t}$.
The projection $p_t$ of $r_t$ should diffuse in 2D:
$$
\langle p_{t+\tau}-p_t \rangle      = 0, \qquad
\langle {(p_{t+\tau}-p_t)}^2\rangle      = 4\, k_B T \mu^S \tau.
$$
Since $P_k$ is the identity in the tangent plane, we used for $\delta B^S_{k,t}$ a vector with $d$ independent components of mean zero, and variance $2\,\tau\mu^S  k_BT$.

\section{Non-deformable set of points (solid)}

We also implemented non-deformable objects called \textit{solids} (see fig.~\ref{fig:objects}) in which the points move together in such a way that the shape and size of the set is conserved.  The number of points $p$ in a \textit{solid}, and their positions $s_i$ can be chosen arbitrarily, and each point is associated with a radius $a_i \geq 0$. The mobility of the \textit{solid} is derived from Stokes's result for the spheres of center $s_i$ and radius $a_i$, neglecting for simplicity the hydrodynamic interactions between the spheres. It is possible to include points with $a_i=0$ provided that  $\sum_i a_i > 0$. In our previous work, we have actually used \textit{solids} where only one $a_i$ was non-zero. These solids moved like isolated spheres, and the points $a_i$ where positions to which forces could be applied.

\subsection{Mobility and Constrained Motion}

Because the set of points should not deform, its elementary motion during a time-step can be written as $(s_i^{t+\tau} - s_i^t ) / \tau= v + \omega \times s_i^t$, where $v$ and $\omega$ are instantaneous translation and rotation speeds.
The spheres of radius $a_i$ in a medium with viscosity $\eta$ have a translational drag coefficient $\xi_i = 6 \pi \eta a_i$, and a rotational drag coefficient $\xi_i^\omega = 8 \pi \eta a_i^3$ \cite{Berg1993}. The forces and torques resulting from the friction of the fluid on the sphere thus read:
\begin{equation}
\tilde{f_i} \;=\; \xi_i \, ( v + \omega\times s_i), \hspace{2cm}
\tilde{M_i} \;=\; \xi_i^\omega \, \omega, \nonumber
\end{equation}
and should match the externally applied forces $f_i$:
\begin{equation}
\sum_i \tilde{f_i} \;=\; \sum_i f_i, \hspace{3cm}
\sum_i s_i \times \tilde{f_i} + \tilde{M_i} \;=\; \sum_i s_i \times f_i \nonumber
\end{equation}
This set of four equations can be solved algebraically in both 2D and 3D, to express $v$ and $\omega$ as a function of the external forces $f_i$. The result always fits in the format of equation~(\ref{eq:dynamics}). It is actually not necessary to calculate the matrix $P$ to run a simulation. It is more efficient to calculate $v$ and $\omega$ when the product $P \mu f$ is needed.
To `reshape' a \textit{solid}, one may restore a reference configuration in the current position and orientation. For this, the best translation and rotation which brings the reference points onto the current points is calculated \cite{Horn1987}. The current points are then replaced by the transformed reference configuration. The Brownian components are calibrated as described before.

%%%%%%%%%%%%%        INTERACTIONS    %%%%%%%%%%%

\section{Interactions between objects}
\label{interactions}

The three objects defined previously can be linked together using elementary interactions.
By adding the contributions of all these interactions in the system, we obtain the linearized force $F(\mathbf{x},t) = A_t\,\mathbf{x} + G_t$, which enters equation (\ref{eq:integration}). In practice, each elementary interaction leads to a small matrix, which needs to be added to the matrix $A_t$ and vector $G_t$, at the right rows and columns to correspond to the appropriate points (see example on figure \ref{fig:interactions}). It is necessary to repeat the procedure at every time step, because the position of the interactions may change with respect to the model-points. We  define four interactions in the case where they connect model-points of the objects. We later explain the procedure to connect intermediate positions between the model-points. This approach can be generalized to more complicated interactions if necessary. For example, it is possible to implement a ring able to slide along a fiber with viscous resistance \cite{Westermann:2006aa}.

\subsection{Connecting an object to a fixed position.}
\label{immobilization}
The simplest way to immobilize an object is to attach a point $a$ within the object to a fixed position $g$. If the stiffness of the link is $k$, the resulting force is $ f_a =  k \, (g - a)$.
In practice, this means adding $-k$ at one diagonal position in matrix $A_t$, and $kg$ to the vector $G_t$ (see fig.~\ref{fig:interactions}).
Such interactions are used to model gliding assays (see fig.~\ref{fig:stability}) in which motors immobilized on a surface propel fibers in solution. Each attached molecular motor leads to an elementary interaction where $g$ corresponds to the place of immobilization, and $a$ corresponds to the position on the fiber at which the motor domain is attached.

\subsection{Connecting two objects.}
\label{connections}
Points from two different objects can be connected by a link of stiffness $k$. The forces between the points are $f_a = -f_b= k \, (b - a)$.
These elementary interactions are effective to model oligomeric motors \cite{Nedelec2002} and more generally any entity able to connect two fibers together (see fig.~\ref{fig:examples}C). In the case of an oligomeric motor, $a$ and $b$ are the positions to which the two motor domains are attached on the fibers.

\subsection{Confinement in a convex shape.}

To confine the objects inside a convex shape, we use a harmonic potential that is flat inside the allowed region, and rises quadratically away from its edge.
Hence, a point $a$ outside the cell volume is subject to a force $f(a)= k (p(a) - a)$, where $p(a)$ is the closest point to $a$ on the edge of the allowed volume. Because $p$ is also the orthogonal projection of $a$, the force corresponds to a friction-less edge. We linearized $f$ as $x \to k \left(\, e_a \cdot (p(a) - x)\,\right) e_a $, where $e_a$ is a unit vector in the direction of $p(a) - a$. 
This linearization corresponds to the tangent plane in $p(a)$, and usually gives a good approximation of $f(a)$ as long as the curvature is small.
To confine a \textit{fiber}, it is sufficient to follow the procedure for its model-points, if the volume is convex, which is the case for example of the cylindrical yeast \textit{S.~pombe} (see fig.~\ref{fig:examples}B). To confine the nucleus of radius $r$ in the same volume, we used a cell volume reduced by $r$. In this way only the center of the \textit{sphere} needs to be tested.

\subsection{Connecting two objects at a given distance.}

A Hookean spring of stiffness $k$ with a non-zero resting length $r$ between two points $a$ and $b$ corresponds to:
$$
f_a = -f_b = -k \left(1 - \frac{r}{|\delta|} \right)\delta,
$$
with $\delta=a-b$. This force should be linearized for $|\delta| \approx r$, leading for $a$ to a term $kr\delta/|\delta|$ in $G_t$ and a contribution in $A_t$  which is:
\begin{equation}
-k\; \frac{ \delta \otimes \delta} { \delta^2 }  \quad\mbox{if}\quad {|\delta| \leq r} \quad\mbox{and}\quad
-k\; \bigg[ I - \frac{r}{|\delta|}[I - \frac{ \delta \otimes \delta} { \delta^2 }] \bigg] \quad\mbox{otherwise},
\end{equation}
and the opposite contributions for $b$. This interaction can be useful to introduce a repulsion between the points. It can for example represent the physical interaction between the nuclear membrane and the microtubules in \textit{S.~pombe} (see fig.~\ref{fig:examples}B).

\subsection{Interpolation of forces}
\label{force:interpolation}
We have discussed connections which were attached to model-points. However, in the case of a \textit{fiber}, a molecule may bind at any position $x$, which is likely to be between two model-points $m_k$ and $m_{k+1}$. When this happens, $a$ is interpolated from the flanking model-points using a coefficient $\alpha =  |m_k x| / |m_k m_{k+1}|$ in $[0,1]$. In the same way, a force $f$ applied in $x$ can be distributed to the model-points as $f_k = (1-\alpha) f$ and $f_{k+1} = \alpha f$. Since this procedure preserves any linearity in the relationship between force and coordinates, the different matrix elements mentioned previously can be used with interpolated points, provided they are multiplied left and right by an appropriate weight matrix.
We can illustrate the procedure for the simplest connection $f_a = -f_b= k \, (b - a)$ of stiffness $k$ between two points $a$ and $b$ (section \ref{connections}), which reads:
$$
\left(\begin{array}{l}
f_a\\
f_b
\end{array}\right)
=
 \left(\begin{array}{rr}
-k & k \\ k & -k
\end{array}\right)
\left(\begin{array}{l}
a\\
b
\end{array}\right).
$$

When $a$ and $b$ are model-points, this $2\times 2$ matrix is a reduction of $A$, corresponding to the $x$, $y$ or $z$- subspaces. This is sufficient in this case because a Hookean spring of null resting length is \textit{isotropic}, that is to say it does not mix $x$, $y$ and $z$ coordinates, and applies similarly to each subspace. This is not the case for all interactions discussed in this section, and it is often necessary to calculate a full matrix. 
Moreover, when $a$ and $b$ are intermediate positions between the model points, we have two indices $k,l$ and two interpolation coefficients $\alpha, \beta$ such that $a = (1-\alpha)\, m_k+\alpha\, m_{k+1}$ and $b = (1-\beta)\, m_l+\beta\, m_{l+1}$. If we define $\overline{\alpha} = 1-\alpha$ and  $\overline{\beta} = 1-\beta$, and
$$w = \left(\begin{array}{rrrr}
\overline{\alpha} & \alpha & 0 & 0 \\
 0  & 0 & \overline{\beta} & \beta\\
\end{array}\right),
$$
we get:
$$
\left(\begin{array}{l} f_k\\ f_{k+1}\\f_l \\ f_{l+1} \end{array}\right)
=
w^t
\left( \begin{array}{l} f_x\\ f_y \end{array}\right)
=
-k \, w^t
\left(\begin{array}{rr} -1 & 1 \\ 1 & -1 \end{array}\right)
w
\left(\begin{array}{l} m_k \\ m_{k+1} \\ m_l \\ m_{l+1} \\ \end{array}\right).
$$

The resulting $4\times4$ matrix is $\overline{w} \, ( -k ) \, \overline{w}^t $, with $\overline{w}^t = ( \overline{\alpha}, \alpha, -\overline{\beta}, -\beta)$. We derive that a matrix made by adding multiple such interactions is symmetric negative-semidefinite ($x^tAx \leq 0$, for any $x$). The fact that this is true for any configuration of the connections guarantees the numerical stability of the method, as explained next.

\section{Numerical Stability and Performance}
\label{stability}

\begin{figure}[p]
 \centering
 \includegraphics[scale=0.8]{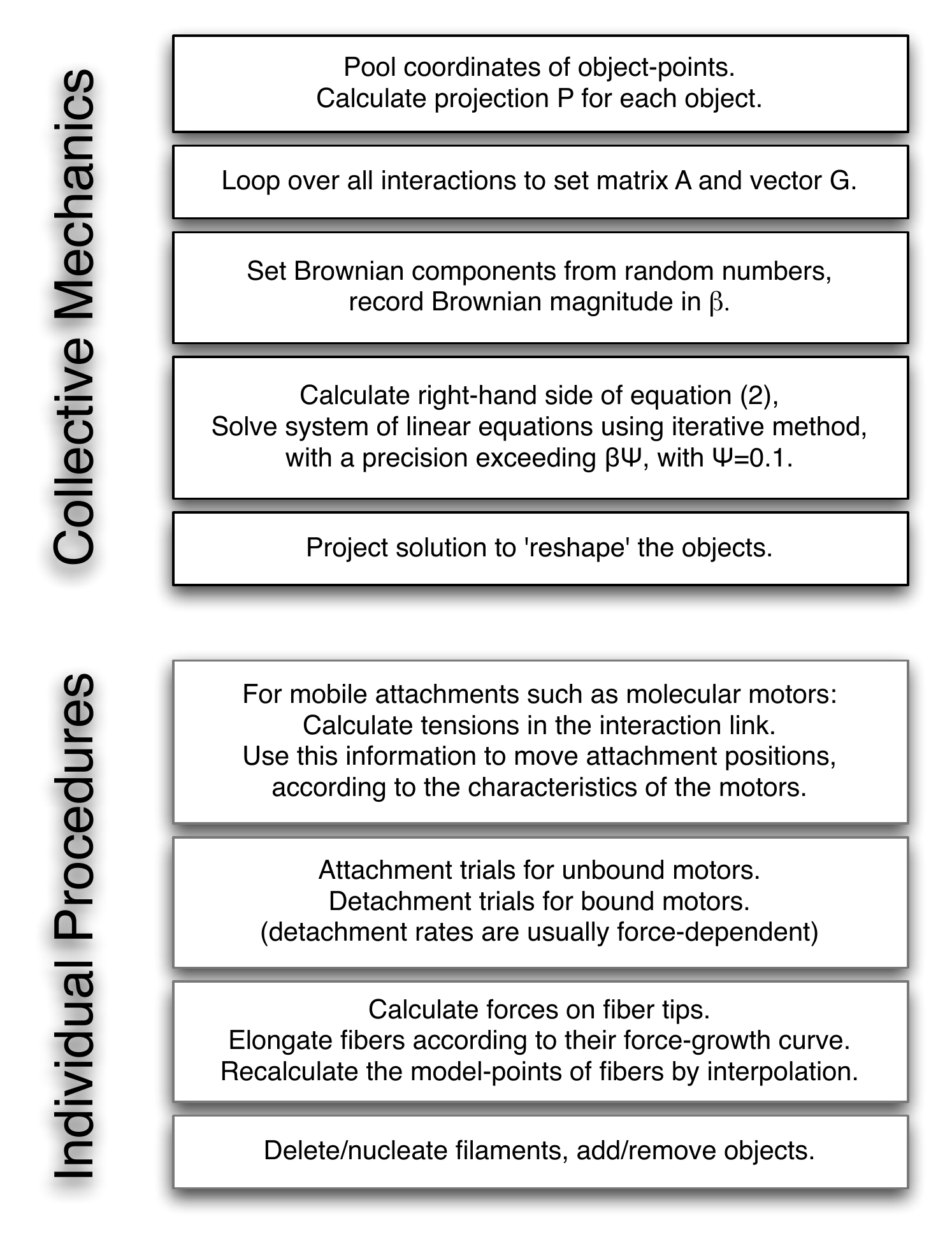}
 \caption{\label{fig:synopsis} \textbf{Synopsis of a simulation time-step.} Sub-steps necessary to simulate a system of molecular motors and dynamic fibers. The collective mechanics corresponds to the algorithm described in the article. As a byproduct of calculating the mechanics, one gets the tensions in the fibers and the forces connecting the fibers. With this information, simulation sub-steps can be performed for the objects independently. Events such as the binding and the unbinding of motors and the nucleation of new filaments will most likely be modeled stochastically.  Depending on the level of details required, less-discrete events may be simulated in a deterministic manner. For example, the active motion of molecular motors and the assembly dynamics of cytoskeletal fibers can be simulated as non-random processes characterized by a force-velocity curve. }
 \end{figure}

We have described all the components of equation (\ref{eq:integration}) which describes the collective mechanics of cellular fibers and other objects.
The necessary steps of the calculation are summarized in figure \ref{fig:synopsis}. 
It is useful at this stage to examine the method mathematically.
This is usually done by looking at two properties: precision and numerical stability \cite{NumericalRecipes}.
The precision is a measure of how the typical error behaves when the time-step $\tau$ becomes small. The numerical stability is a measure of how large $\tau$ can be, before the calculation fails.
Numerical precision is important for deterministic equations, for example to predict the trajectories of celestial bodies. However, this is not so critical at the cellular scale.
In fact, to simulate the Brownian motion present in the cell, a random term $\delta B \sim \sqrt{\tau}$ was included in equation (\ref{eq:integration}). The presence of this `noise' indicates that the physics itself limits the precision at which the position of an object can be predicted. This fact undermines the usefulness of high precision schemes. The implicit method that we have described is of order one: the step's error scales like $O(\tau^2)$, which is better than the physical `noise' in $\sqrt{\tau}$. We found that it was not practically useful to use higher order numerical schemes. 

\begin{figure}[p]
 \centering
 \includegraphics{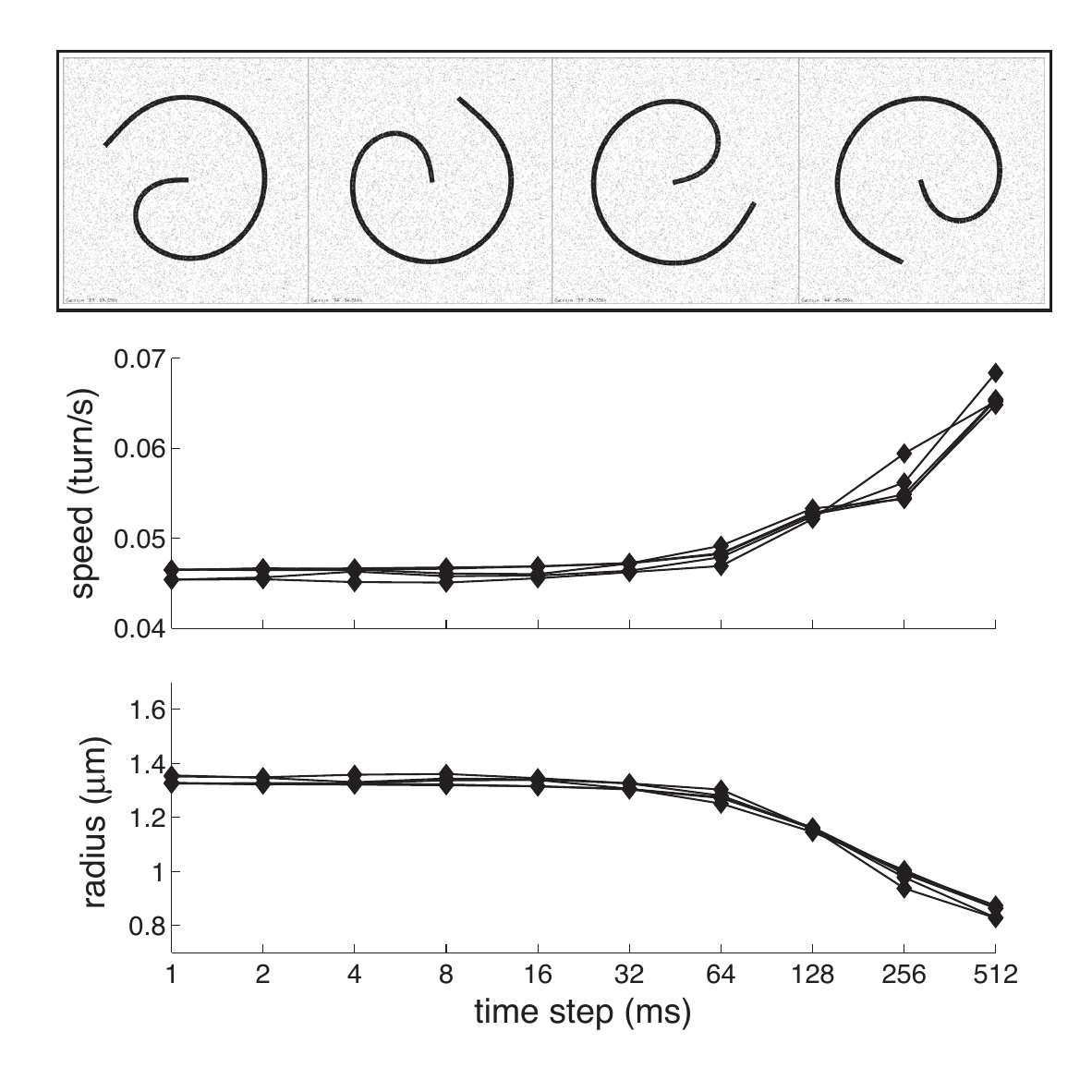}
 \caption{\label{fig:stability} \textbf{Numerical stability of the integration scheme.} {\bf Top:} A gliding assay where a filament is attached at its end (time-intervals of 5s). The motors pushing the fiber lead to the formation of a rotating spiral, as observed experimentally \cite{Bourdieu1995}. The rotation speed and maximum radius of the spiral can be calculated from the parameters of the system: 16000 motors cover an area of $2\times2 \um$, and have the characteristics of kinesin: stall force $f_{\mbox{\scriptsize max}} = 5\,pN$, unloaded speed $0.4 \um/s$, binding rate $10\,s^{-1}$, unbinding rate $p_{\mbox{\scriptsize off}} = 0.5\, s^{-1} \exp(\mbox{force}/2.5\,pN)$, maximum binding distance $10\, \nm$ and stiffness $200\, pN/\um$. The microtubule of length $8$ \um\ has a rigidity of $20\, pN \um^2$. It is constrained at the minus end by a link of stiffness $4000\, pN/\um$. The effective viscosity is $0.02\, pN\,s\, \um^{-2}$. 
 \textbf{Bottom:} The configuration is simulated for different values of the time-step $\tau$, with accurate results for $\tau < 20$ ms. The algorithm is numerically stable, and even produces a spiral with $\tau \sim0.5$ s. However, the radius is then under-estimated, and the rotation speed overestimated. Another critical parameter, the distance $\rho$ between the points on the fiber was also varied. The results shown for $\rho = 0.1$, $0.2$, $0.4$ and $0.5$ \um\ (different lines) are similar, because all these values are appropriate. The calculations were inaccurate however with $\rho = 0.8$ \um\ (data not shown). This is expected considering that the radius of the spiral is $\sim 1.4$\um.}
\end{figure}

In contrast, the numerical stability of the method is most important. Indeed, explicit schemes usually converge only if the time-step is small. In general, a condition like $\tau \mu k < 1$ must be fulfilled, where $\mu$ is the mobility of a point in the system, and $k$ the stiffness of the interaction potential. For example, we looked at a test-case in which a microtubule is pushed by immobilized motors (see \cite{Bourdieu1995} and Fig.~\ref{fig:stability}). It can be simulated explicitly only if $\tau < 1\, \mu$s, but the implicit method can use larger time-steps. To achieve this stability, we treated the repulsive and attractive interactions in the system differently. Compressive forces in the fibers (which are repulsive in nature) were replaced by constraints. All the other forces were attractive. This ensured that $A_t$ would be negative-semidefinite (this result was proven in section \ref{connections} for Hookean interactions of null resting length). Mathematically, because $P_t$ is an orthogonal projection, we can show that the eigenvalues of $I - \tau\mu P_t A_t$ are always greater than $1$, for any value of $\tau$. This implies that our integration scheme is unconditionally stable.
For the other elementary interactions, some instabilities may appear, but only for very high values of the time step (not shown).

Beyond stability, other considerations naturally limit the choice of $\tau$. In particular the iterative solver might not converge when $\tau$ is large. The optimal time-step generally depends on the problem studied, and it is best to perform systematic trials to find it. For the test-case (see fig.~\ref{fig:stability}), the results are consistent for $\tau < 20$ ms. This means that a value of $5$ or $10$ ms would be appropriate.
The computational requirements depend on the total number of steps (total time/time-step), but also on the cost of individual steps. An implicit step of integration is always more costly than an explicit step, because a linear system must be solved. However, the use of sparse matrix techniques reduces the additional work. In practice the considerable reduction in the number of steps makes implicit simulations faster (in the test-case, this gain is $10^4$, using $\tau = 10$ ms instead of $1\, \mu$s). Increasing the execution speed is essential if many simulations need to be performed.
Implicit methods require increased numerical labor, of which we have illustrated the main difficulties.
Using the method described here, we can simulate the examples shown in figure~2 B, C \& D much faster than real time using one processor (www.cytosim.org).

\section{Other Elements of a Cytoskeletal Simulation}

In addition to mechanics, a cytoskeletal simulation such as cyto\textbf{sim} must include additional aspects such as the motion of molecular motors, their binding/unbinding dynamics, as well as the transitions between growth and shrinkage of dynamic fibers.
These processes can be modeled most simply by executing small sub-routines after the Brownian mechanics has been calculated, because they correspond to independent operations (see fig.~\ref{fig:synopsis}). 
However, two particularly important aspects of cytoskeletal physics need to be mentioned.
Firstly, only in very particular cases can we approximate the system as a well-mixed reactor. At least some of the molecules should be spatially resolved. 
Secondly, the mechanics commonly affects the chemistry. For instance the rates of certain key reactions are force-dependent. This is the case for the unbinding rates of molecular motors and for their stepping rate (see below).
Because these elements are essential for modeling the system accurately, it will rarely be possible to apply algorithms developed for purely chemical systems (\textit{eg.}~the Gillespie algorithms \cite{Gillespie:2007aa} or even spatially resolved methods \cite{Hattne:2005aa}) without extensive modifications. We can however use simple and robust simulation strategies, as illustrated below in the case of molecular motors.

\subsection{Modeling Molecular Motors}

\begin{figure}[p]
\centering
\includegraphics[scale=0.8]{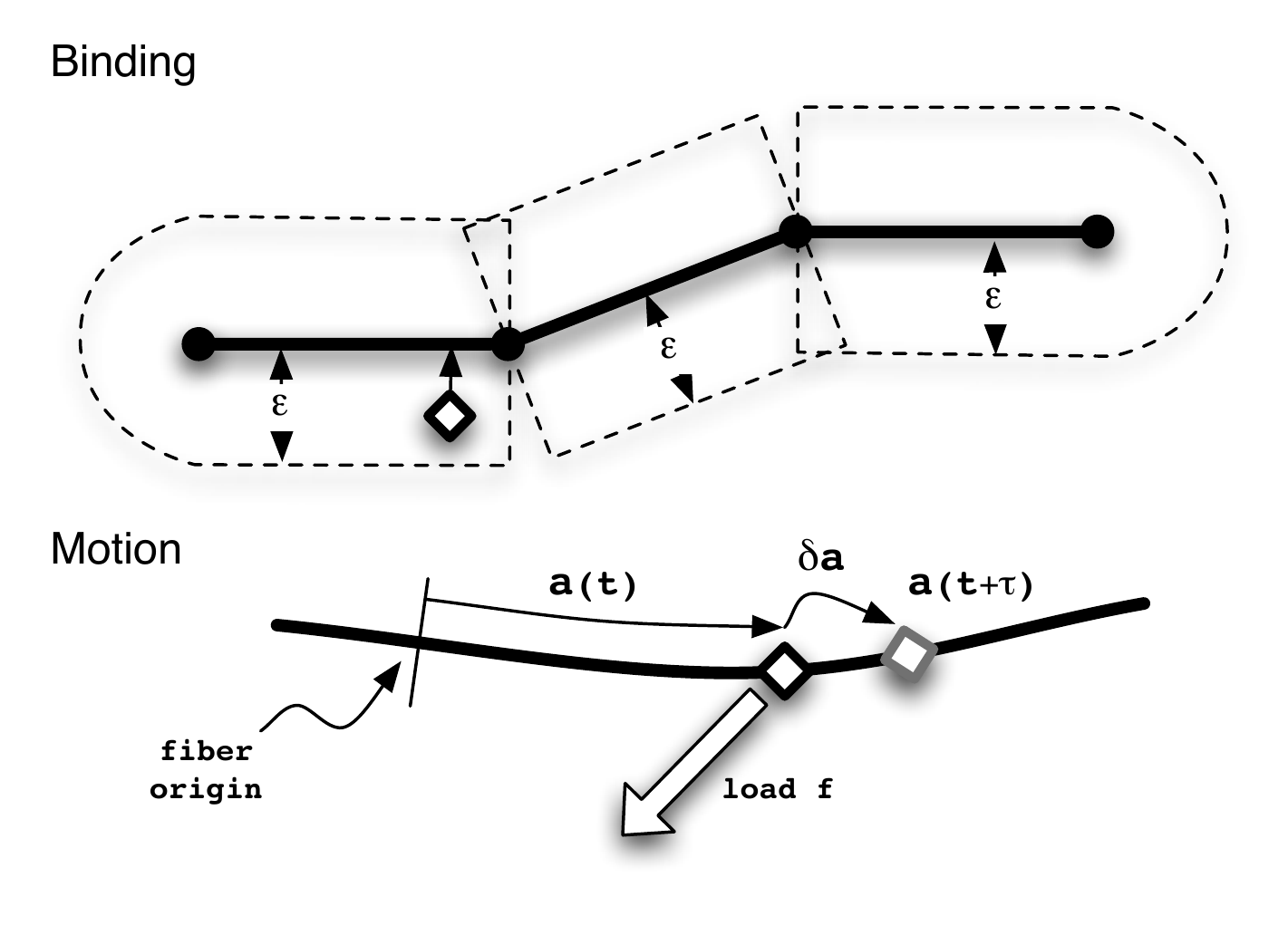}
 \caption{\label{fig:motors} \textbf{Molecular Motors.}  \textbf{Top:} An unbound motor (diamond) is represented by a position. Attachment occurs on the closest site on the fiber-segment, provided this site is within a distance $\epsilon$ (dashed lines). The capture regions of the segments are truncated such that they cover exactly the region located at a distance $\epsilon$ from a straight fiber. When the fiber is not straight, the gaps and overlaps exactly compensate each other. \textbf{Bottom:} A bound motor is represented by a pointer to a fiber, and by a curvilinear abscissa $a(t)$ measured from a fixed origin on the fiber. This defines the position of the motor along the fiber independently of the mathematical representation of the fiber. The motor sub-model needs to decide whether the motor should detach during the interval of time $\tau$, or it needs to calculate the displacement $\delta a$ during the same interval. For this, it can use the load $f$ calculated during the collective mechanics, and other properties associated with the fiber, such as the proximity of the ends, or information on the crowdedness of the binding sites on the fiber. }
 \end{figure}

In cyto\textbf{sim}, a motor is characterized by a position, when it is not attached, and by a pointer to a fiber and a curvilinear abscissa, when it is attached (see fig.~\ref{fig:motors}). The abscissa is the distance, measured along the fiber, between a reference and the attachment position. It is necessary to use a reference which is fixed with respect to the physical lattice, because the model-points of a \textit{fiber} are themselves updated as the fiber grows (see fig.~\ref{fig:fibers}).
This description neatly separates the details of how the mechanics is implemented from the routines simulating the motors \textit{per se}. 
This means that the interface with the rest of the program can be very simple, with only two procedures: \textit{step(f)} and \textit{attach(m)}. 

\subsubsection{Active Motion.}

The first procedure \textit{step(f)} simulates the possible actions of a bound motor. The argument $f$ is the load of the motor calculated during the collective mechanics. 
The procedure should decide to detach the motor, or to update the abscissa $a$ according to a microscopic model for the interval $\tau$.
For a well characterized motor like kinesin, a classical model is based on the measured characteristics of the motion: the abscissa is increased by $\delta a = \tau v_{\mbox{\scriptsize max}} ( 1 - f/f_{\mbox{\scriptsize stall}} )$. In addition, a force-dependent unbinding rate $p_{\mbox{\scriptsize off}} = p_0 \exp(|f|/f_0)$ is used to model the dissociation from the fiber.  $v_{\mbox{\scriptsize max}}$, $p_0$, $f_0$ and $f_{\mbox{\scriptsize stall}}$ are characteristics of the motor that have been measured for kinesin \cite{Howard2001}.
With this model, the fibers are continuous tracks along which motors may be located anywhere.
Alternatively, we may model the motion of a motor as a succession of discrete stochastic steps.
In this case, the motor does one of four things: stay immobile, detach, take a step toward the minus-end or take a step toward the plus-end. This means that if the motor does not detach, the abscissa is either unchanged, or it is increased or decreased by the step size (8 \nm).  The procedure \textit{step(f)} calculates the probabilities of these events as a function of the force $f$ for the interval $\tau$, and selects one of them. This model is quite attractive, because these probabilities are actually available for kinesin \cite{Carter:2005aa}.
Most models describing the movement of motors \cite{Kolomeisky:2007aa} can be summarized similarly with a function  \textit{step(f)}.

\subsubsection{Attachment to Fibers.}

\begin{figure}[p]
 \centering
 \includegraphics[scale=0.8]{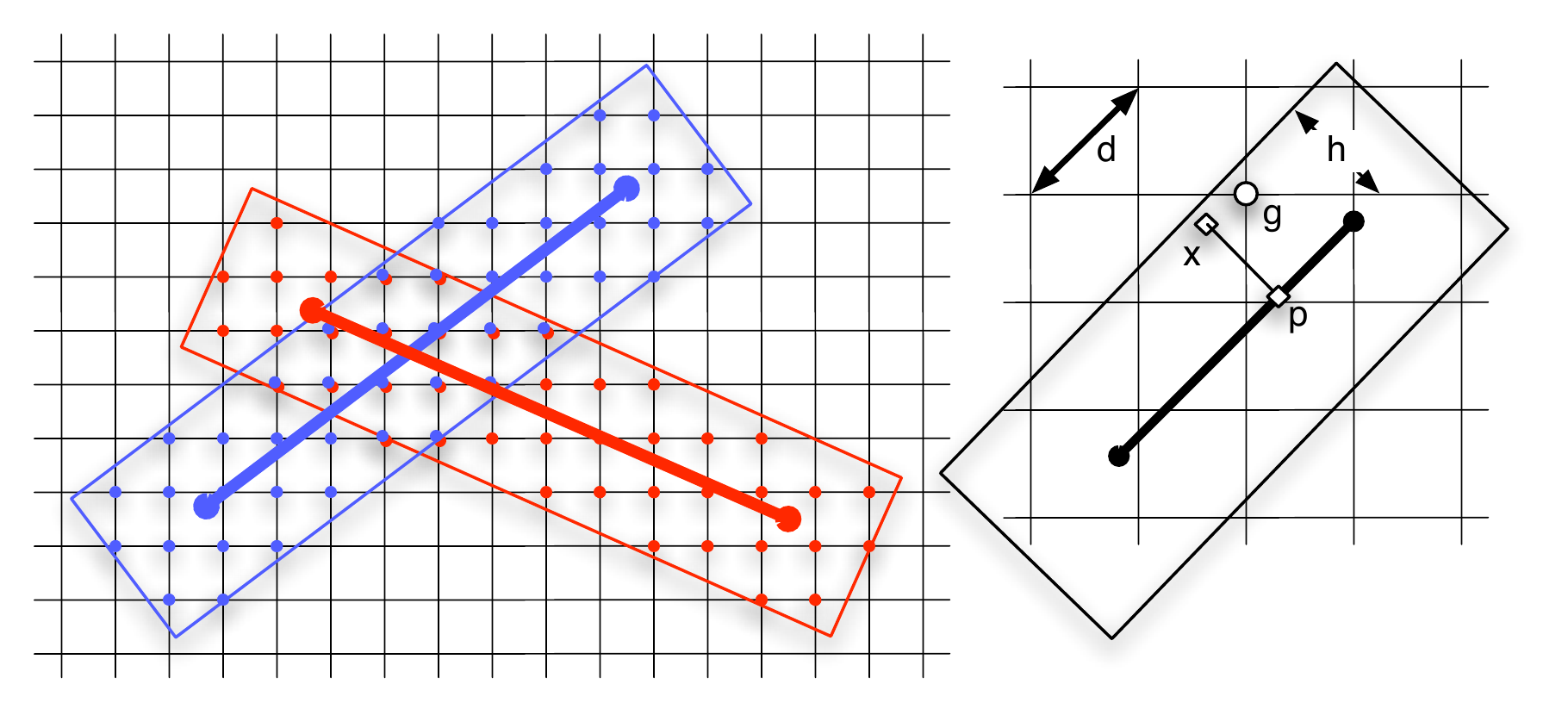}
 \caption{\label{fig:divide} \textbf{Divide and Conquer algorithm.} To simulate the attachments to fibers, we must be able to find all the fiber-segments which are within a distance $\epsilon$ from an arbitrary position $x$. We can proceed according to the following two steps method: \textbf{Divide (left):} A grid is set in space, each node of the grid being associated with a list of segments. The segments are recorded on the grid, at the nodes located at a distance $h$ or less ($h$ will be defined later). This operation is performed in 2D using standard rasterizer codes derived from computer graphics, which are optimized to scan all points with integer coordinates located inside an arbitrary polygon. We rasterize the rectangles built around the segments at a distance $h$. For example, on this diagram, the blue segment is recorded at the blue points, and the red segment at the red points. 
In 3D, the  rectangular volume can be rasterized following the same principles as in 2D. 
\textbf{Conquer (right):} After the segments have been distributed over the grid, one can quickly find which ones are near $x$: one needs to check only the segments recorded at the grid point $g$ closest to $x$.  One will find all segments located at distance $h - d/2$ or less from $x$, since $|gx| < d/2$, where $d$ is the diagonal of the grid. Hence to find all the segments closer than $\epsilon$, one sets $h=\epsilon + d/2$ during the rasterizing operation. 
\textbf{Note:} 
The grid does not need to be square (the unit cell can be rectangular) and it can be adjusted for optimal performance. If the grid is too fine, it will use a lot of memory, and rasterizing will be slow. If the grid is coarse ($d$ large), the number of candidates returned for a point $x$ will be larger.  Experimentation may be necessary to optimize the grid, but the procedure provides exact results for any cell size.  }
 \end{figure}

The second procedure necessary to model motors, \textit{attach(m)} simply decides if a unbound motor binds or not to a site $m$. 
Usually the model would specify $\epsilon$, a maximum distance at which a motor may bind from its current position (see fig.~\ref{fig:motors}). In addition, the molecule would bind at the closest site on the fiber (the orthogonal projection) with a certain molecular binding rate $k_{on}$ ($s^{-1}$).
To simulate attachments, one therefore needs to first find the fiber-segments which are closer than $\epsilon$, typically from all the positions $x$ at which molecular motors are located. 
For each point $x$, the list of candidates should then be shuffled, to ensure a random ordering of the segments.
The molecular binding rate can finally be tested sequentially for each segment in the list, for example by comparing $\tau k_{on}$ with a random number $\theta$ in [0,1]. The first successful trial is followed by attachment.
If done naively, the first step of the operation may require calculating the distance of all points to all fiber-segments, and thus a great deal of computation for many motors.  
To avoid this bottleneck in cyto{\bf sim}, a divide and conquer algorithm was developed (see fig.~\ref{fig:divide}).
Its goal is to limit the number of segments that need to be tested to find those which are close to $x$. The geometrical distance between $x$ and these segments is calculated using the vector cross-product to exactly determine which ones are closer than $\epsilon$.
Reducing the number of tested segments is sufficient to accelerate the simulation.

\section{Conclusion}

The method described here is efficient to simulate sparsely connected networks of filaments. It applies to many \textit{in vivo} situations, because the connections between fibers are usually mediated by proteins that are small compared to the fibers, and consequently the fibers are only locally connected. We have modeled fibers as oriented lines, which is sufficient to calculate the extent of bending. It may be necessary in the future to include more details such as writhe, since cytoskeletal fibers also have a torsional rigidity.  The method can be extended in several other ways. One could for instance easily model discrete binding sites on the fibers. This may be important if the fibers are highly covered and molecules compete or interact while bound to the lattice.
It is also possible to extend the overdamped mechanics by adding hydrodynamic effects. It will be very exciting to integrate fiber mechanics with membrane dynamics, since membranes and cytoskeleton contribute synergistically to cellular architecture, but this might take some time.
Cellular chemistry, reaction-diffusion of the components in the cell, gene expression networks, can be added more simply.
This can be done by interfacing our software with other tools (eg.~the Virtual Cell project), which already cover some of these aspects of physiology. We did not discuss here implementation issues, but the scale of the task should remind us of their importance.
Software modularity is essential to divide the development effort in separate projects of manageable size. Sub-models or algorithms should be developed and tested separately, in such a way that they can be added or removed from the integrative software easily.
Dividing the work among different groups is the best way to produce the high-quality cellular simulations that biology needs.

\ack
This method was designed in 2001 \cite{Nedelec2002} and extended by Dietrich Foethke to \textit{spheres}. We thank the members of our laboratory, and in particular Rose Loughlin and Cleopatra Kozlowski for their help in developing Cytosim, and for critically reading this manuscript. We thank Jonathan Ward for his critical reading, Tony Lelievre and Rafe Mazzeo for mathematical insights. We acknowledge support from IBM, BioMS (www.bioms.de), the Volkswagenstifftung initiative ``new conceptual approaches to modeling and simulation of complex systems'' and HFSP grant RGY84.

  % BIBLIOGRAPHY   
\clearpage
\bibliography{cytosim_math}
\bibliographystyle{unsrt}

\end{document}